%% file: main.tex
\documentclass[a4paper,11pt]{article}
\usepackage{jcappub} 
\usepackage{lineno}
\usepackage{cleveref}
\usepackage[squaren]{SIunits}

\arxivnumber{1234.56789} 
\title{A cross-calibration approach for polarisation-sensitive detectors in CMB experiments: application to LiteBIRD’s polarisation angle calibration}

\input{authors_affiliations_LB_2025.08_general_jcap}

\emailAdd{alessandro.novelli@unimib.it}

\abstract{We present a cross-calibration algorithm that determines the relative calibration of detectors within the same frequency band of a CMB experiment. The method iteratively compares single-detector maps with frequency-band maps, and it can be applied to any calibration parameter that can be corrected through reprocessing of the instrument's data at the map level. This allows for validation and correction of calibration both before and after observations.\\
We validate the pipeline by calibrating the polarisation angle of simulated sky maps from the upcoming \textit{LiteBIRD} experiment. The cross-calibration algorithm has been tested both in the case of random miscalibration of each individual detector and in the case of rotation of entire detector wafers. In both cases, the cross-calibration algorithm converges to the correct value and we are able to identify and correct relative miscalibrations in the polarisation angle of each detector with arcminute precision. Absolute calibration can then be addressed either through ground-based measurements, or through the $E$-$B$ nulling technique.\\
To assess compliance with \textit{LiteBIRD}'s systematics budget, we propagate the residual miscalibration uncertainties through component separation and tensor-to-scalar ratio ($r$) estimation pipelines using both parametric (\texttt{FgBuster}) and blind (\texttt{HILC}) methods. In both cases, the induced bias on $r$ remains well below the \textit{LiteBIRD} maximum budget for systematics of $\delta r < 6.5\times10^{-6}$. This confirms that the cross-calibration method presented is suited for use in modern high-precision 
CMB experiments.}

\begin{document}
\maketitle
\flushbottom

\section{Introduction}
\label{sec: Introduction}
In the last 50 years orbital and sub-orbital experiments \cite{PenziasWilson, COBE, Boomerang} have measured the CMB temperature anisotropies to nearly the cosmic variance limit, constrained cosmological parameters, and expanded our knowledge regarding the evolution of the Universe. One of the current challenges in observational cosmology consists of improving the accuracy of current polarisation maps of the microwave sky \cite{WMAP, Planck2018, QUIJOTE} and detecting the primordial $B$ modes. Such an achievement would allow us to determine the tensor-to-scalar ratio $r$ and the energy scale of cosmic inflation. Unfortunately, the $B$-mode signal is expected to be extremely faint and is contaminated by Galactic foreground emission. Therefore, to achieve a detection, any instrument must rely on a large number of well-characterised detectors at several frequency bands. The presence of unwanted instrumental systematic effects and miscalibrations is considered one of the critical issues in realizing this achievement. Current observational constraints are set at $r < 0.032$ (95\% CL) by \cite{Tristram_r, Galloni_r}.\\
The purpose of this paper is to present a cross-calibration approach that performs relative calibration of detectors within each frequency channel of a CMB instrument. The methodology can be applied to any detector parameter that can be corrected through reprocessing of the instrument’s data. Examples of such correctable systematics include gain, polarisation efficiency, and polarisation angle, all of which can be adjusted at the map level if a miscalibration is identified. In contrast, this cross-calibration method is not applicable to systematic effects such as cosmic ray glitches or additional noise sources, where the effects cannot be deterministically corrected at the map level. To demonstrate the capabilities of the proposed cross-calibration algorithm we will use it to calibrate the polarisation angle of the detectors of the \textit{LiteBIRD} satellite.\\
\textit{LiteBIRD} (The Lite (Light) spacecraft for the study of $B$-mode polarisation and Inflation from cosmic background Radiation Detection) \cite{PTEP} is a space-borne mission selected by JAXA to measure the polarisation of the CMB with unprecedented accuracy. Its goal is to determine the tensor-to-scalar ratio with a precision of $\sigma_r\leq 10^{-3}$ and it aims to do so by observing the sky with around 5000 polarisation sensitive bolometers distributed in 22 channels across 15 different frequency bands. The launch is projected for the 2030s and observations will last 3 years from the Earth-Sun Lagrange point L2.\\
The next sections are organised as follows. \Cref{subsec: Simulated Sky-Maps} describes how we used \texttt{litebird\_sim} \cite{litebrid_sim} to obtain synthetic sky maps as observed by the \textit{LiteBIRD} instrument. In \cref{subsec: Relative miscalibrations} we describe how we model a miscalibration of the polarisation angle $\theta_\text{pol}$ of the detectors and the effect this has on the data. \Cref{subsec: Cross-calibration} describes the cross-calibration algorithm. \Cref{subsec: Residual miscalib} describes how we can assess the impact on $r$ of the residual miscalibrations after cross-calibration. Finally we present the results obtained in \cref{sec: Results} and we provide final considerations in \cref{sec: Conclusions}.
\section{Simulations}
\subsection{Simulated Sky-Maps}
\label{subsec: Simulated Sky-Maps}
We consider a sky model composed of CMB and Galactic foreground emission. CMB maps have been generated starting from the power spectra computed with the \texttt{CAMB} package \cite{CAMB} using the \textit{Planck} best-fit $\Lambda$CDM cosmological parameters \cite{Planck_PowerSpectra}: $H_0 = 67.5\,\mathrm{km\,s^{-1}\,Mpc^{-1}}$, $\Omega_\mathrm{b}h^2 = 0.022$, $\Omega_\mathrm{c}h^2 = 0.122$, $A_\mathrm{s} = 2 \times 10^{-9}$, $n_\mathrm{s}=0.965$, $\tau=0.06$, $m_\nu = 0.06\,\mathrm{eV}$, $\Omega_k = 0$, and $r=0$.\\
Since we want to calibrate the polarisation angle of the detectors, we are mainly interested in foregrounds that have a polarised emission. For this reason, we model the Galactic emission as composed of dust and synchrotron emission. We are neglecting other sources of Galactic emission that are unpolarised or have a low polarisation fraction such as free-free radiation, Anomalous Microwave Emission (AME) and CO lines. Polarised dust emission is dominant at high frequencies $\nu \gtrsim 90 \,\giga \hertz$ and is usually modelled with a modified black-body spectrum \cite{Planck_Dust}:
\begin{equation}
[I,Q,U]_\text{dust}=A_{[I,Q,U]_{\text{dust}}}\left(\frac{\nu}{\nu_\text{ref}}\right)^{\beta_\text{d}-2}\frac{B(\nu, T_\text{d})}{B(\nu_\text{ref}, T_\text{d})}
\end{equation}
where $\beta_\text{d}$ is the dust spectral index, $B(\nu, T)$ is the black-body spectrum, $T_\text{d}$ is the dust temperature and $\nu_\text{ref}$ the reference frequency used to define the templates $A_{[I,Q,U]_{\text{dust}}}$.\\ 
Synchrotron emission is modelled using a power-law spectrum \cite{WMAP_synchrotron}:
\begin{equation}
    [I,Q,U]_\text{sync}=A_{[I,Q,U]_\text{sync}}\left(\frac{\nu}{\nu_\text{ref}}\right)^{\beta_\text{s}-2}
\end{equation}
where $\beta_\text{s}$ represents the spectral index of the synchrotron and $\nu_\text{ref}$ is the reference frequency used to define the templates $A_{[I,Q,U]_\text{sync}}$.\\
For the purpose of this analysis, we used the \texttt{d0} and \texttt{s0} models from \texttt{PySM}\footnote{https://pysm3.readthedocs.io/} to represent the dust and synchrotron emissions. These models are characterised by uniform spectral parameters over the whole sky $\beta_\text{s}=-3$, $\beta_\text{d}=1.54$, and $T_\text{d}=20\,\kelvin$. The choice to adopt these simplified models does not affect the results of the cross-calibration algorithm and significantly reduces the computational cost of the analysis reported in \cref{subsec: Residual miscalib}.\\
In its original configuration \textit{LiteBIRD} \cite{PTEP} is a satellite composed of three instruments: the Low-Frequency Telescope (LFT) operates in the range $31\,\giga \hertz \leq \nu\leq 161\, \giga \hertz$, while the Medium-Frequency and High-Frequency telescopes cover the $89\, \giga\hertz \leq\nu \leq 224\, \giga\hertz$ and $166\,\giga\hertz\leq \nu\leq448\, \giga \hertz$ frequency ranges respectively. The detectors of the instruments have an angular resolution ranging from $24\, \text{arcmin}$ to $71 \,\text{arcmin}$. The first optical element of each instrument is a rotating Half-Wave Plate (HWP) polarisation modulator, which reduces the presence of the $1/f$ noise in polarisation maps as well as mitigates some systematic effects. The instrument properties used for the analysis of this paper have been taken from \cite{PTEP} and are reported in \cref{tab: LiteBIRD table}.\\
To produce simulated maps of the \textit{LiteBIRD} instrument we use \texttt{litebird\_sim} \cite{litebrid_sim} to sum a realisation of the CMB with foreground from \texttt{PySM} and smooth the maps to the angular resolution of each individual band. Single-detector maps are produced using realistic per-detector hitmaps obtained from the full \textit{LiteBIRD} scanning strategy simulation via \texttt{litebird\_sim}, which account for the actual pixel coverage of each detector. We then add to the single-detector maps realisations of instrumental noise, which, under the assumption of an ideal HWP, is taken to be white and uncorrelated between detectors. Finally, we deconvolve each map with their respective beam and re-convolve it with the lowest \textit{LiteBIRD} resolution of $70.5\,\text{arcmin}$. The result of these simulations is a pair of $Q^\text{sky}_{\nu,\text{det}}$ and $U^\text{sky}_{\nu,\text{det}}$  maps of the sky as observed by each detector of \textit{LiteBIRD} and then brought to a common angular resolution of $70.5\,\text{arcmin}$.
\begin{table}[t]
\centering
\begin{tabular}{llllll}
\hline
\textbf{Telescope} & \textbf{Freq.} & \textbf{Channel label} & \textbf{FWHM} & \textbf{Sensitivity } & \textbf{N$_\text{bol}$} \\
&  [GHz] & & [arcmin] & [$\mu$K-arcmin] & \\
\hline
LFT & 40  & LFT-40   & 70.5 & 37.42 & 48  \\
LFT & 50  & LFT-50   & 58.5 & 33.46 & 24  \\
LFT & 60  & LFT-60   & 51.1 & 21.31 & 48  \\
LFT & 68  & LFT-68a  & 41.6 & 19.91 & 144 \\
LFT & 68  & LFT-68b  & 47.1 & 31.77 & 24  \\
LFT & 78  & LFT-78a  & 36.9 & 15.55 & 144 \\
LFT & 78  & LFT-78b  & 43.8 & 19.13 & 48  \\
LFT & 89  & LFT-89a  & 33.0 & 12.28 & 144 \\
LFT & 89  & LFT-89b  & 41.5 & 28.77 & 24  \\
LFT & 100 & LFT-100  & 30.2 & 10.34 & 144 \\
LFT & 119 & LFT-119  & 26.3 & 7.69  & 144 \\
LFT & 140 & LFT-140  & 23.7 & 7.25  & 144 \\
MFT & 100 & MFT-100  & 37.8 & 8.48  & 366 \\
MFT & 119 & MFT-119  & 33.6 & 5.70  & 488 \\
MFT & 140 & MFT-140  & 30.8 & 6.38  & 366 \\
MFT & 166 & MFT-166  & 28.9 & 5.57  & 488 \\
MFT & 195 & MFT-195  & 28.0 & 7.05  & 366 \\
HFT & 195 & HFT-195  & 28.6 & 10.50 & 254 \\
HFT & 235 & HFT-235  & 24.7 & 10.79 & 254 \\
HFT & 280 & HFT-280  & 22.5 & 13.80 & 254 \\
HFT & 337 & HFT-337  & 20.9 & 21.95 & 254 \\
HFT & 402 & HFT-402  & 17.9 & 47.45 & 338 \\
\hline
\end{tabular}
\caption{Specifications of the \textit{LiteBIRD} frequency bands \cite{PTEP}. From left to right: the telescope, the operating frequency, the band label, the beamsize of the detectors, the polarisation sensitivity, and the number of bolometers.}
\label{tab: LiteBIRD table}
\end{table}
\subsection{Relative miscalibrations of the absolute polarization angle}
\label{subsec: Relative miscalibrations}
For the purpose of this paper, we will consider the uncertainty on the polarisation angle of the detectors $\theta_\text{pol}$ to be the leading systematic effect of our instrument. We introduce the polarisation angle of each detector as a mixing term between the Q and U maps that only applies to the sky emission and not the noise:
\begin{equation}
m^{\text{obs}}_{\nu,\text{det}}=Q^{\text{obs}}_{\nu,\text{det}}+iU^{\text{obs}}_{\nu,\text{det}}=\left(Q^\text{sky}_{\nu,\text{det}}+iU^\text{sky}_{\nu,\text{det}}\right)\,e^{-2i\theta_\text{pol}^{\nu, \text{det}}}+n\text{ .}
\end{equation}
Here $Q^\text{sky}_{\nu,\text{det}}$ and $U^\text{sky}_{\nu,\text{det}}$ represent the Stokes parameters of the polarised sky as simulated in \cref{subsec: Simulated Sky-Maps}, while $m^{\text{obs}}_{\nu,\text{det}}$ is the sky map measured by a given detector of frequency band $\nu$. The polarisation angle $\theta_\text{pol}^{\nu, \text{det}}$ of each detector is assumed to be homogeneous and constant over time.\\
In the case of a perfect calibration, the orientation of each detector is perfectly determined and the maps of the detectors can all be brought to a common sky orientation. If we consider the maps $[Q,U]^\text{sky}_{\nu,\text{det}}$ to have already been brought to the same orientation in the case of a perfect calibration we have: $\theta_\text{pol}^{\nu, \text{det}}=0$, for every frequency and detector. To simulate calibration errors, we can generate for each detector a random polarisation angle extracted from a Gaussian distribution with standard deviation $\sigma(\theta_\text{pol}^\nu)$:
\begin{equation}
    \theta_\text{pol}^{\nu, \text{det}}=0+\mathcal{N}\left(0,\sigma(\theta_\text{pol}^{\nu})\right)
    \label{eq: theta_pol realisation}
\end{equation}
Since the cross-calibration algorithm is purely relative by construction, it cannot determine the mean polarisation angle $\langle \theta_\text{pol}^{\nu, \text{det}}\rangle_{\text{det}}$ of any realisation of the detectors within a channel, which must be addressed by absolute calibration. Therefore, without loss of generality, in our simulations we impose $\langle \theta_\text{pol}^{\nu, \text{det}}\rangle_{\text{det}}=0$ within each frequency band, so as to isolate only the detector-to-detector scatter that the cross-calibration algorithm is designed to correct.\\
Throughout the remainder of this paper we will use $\theta_\text{pol}^{\nu, \text{det}}$ to identify the miscalibration of a certain detector and $\sigma(\theta_\text{pol}^{\nu})$ to identify the uncertainty on the polarisation angle of the detectors of frequency band $\nu$.
\subsection{Cross-calibration algorithm}
\label{subsec: Cross-calibration}
We now present a cross calibration algorithm that performs the relative calibration of the detectors within the same frequency channel of a CMB instrument. We introduce the pipeline in the context of the relative calibration of the polarisation angle of the detectors of the \textit{LiteBIRD} satellite; however, as mentioned in \cref{sec: Introduction}, it is applicable to detector miscalibrations that can be modeled and corrected through reprocessing of the instrument’s data at the map level.\\
The cross-calibration algorithm is as follows:
\begin{enumerate}
    \item Retrieve individual detector maps $m_{\nu,\text{det}}^{\text{obs}}$, each affected by an unknown polarisation angle miscalibration $\theta_\text{pol}^{\nu,\text{det}}$.
    \item Combine single-detector maps into channel-wide maps. This was done by averaging the detector maps of the same channel weighted by their per-pixel hitmap counts $h_{\nu,\text{det}}(p)$ in pixel space: $m_{\nu}^{\text{obs}}(p) = \frac{\sum_\text{det} h_{\nu,\text{det}}(p)\, 
    m_{\nu,\text{det}}^{\text{obs}}(p)}{\sum_\text{det} 
    h_{\nu,\text{det}}(p)}$.
    \item 
    For each detector, determine the value of $\tilde{\theta}_\text{pol}^{\nu,\text{det}}$ that best aligns its map with the frequency-averaged one.
    This can be obtained minimising the quantity:
    \begin{equation}
        \chi^2_\text{det}(m_{\nu,\text{det}}^{\text{obs}}, \tilde{\theta}_\text{pol}^{\nu,\text{det}})=\sum_{p=1}^{N_\text{pixel}}\frac{\left(m_{\nu}^{\text{obs}} - m_{\nu,\text{det}}^{\text{obs}} e^{2i\tilde{\theta}_\text{pol}^{\nu,\text{det}}}\right)^2}{\sigma(m_{\nu,\text{det}}^{\text{obs}})^2}
        \label{eq: chi_squared}
    \end{equation}
    where $\sigma$ denotes the noise uncertainty in each pixel. 
    \item Rotate all detector maps by their estimated angle to correct the miscalibration: $m_{\nu,\text{det}}^{\text{corr}}= m_{\nu,\text{det}}^{\text{obs}}\,e^{2i\tilde{\theta}_\text{pol}^{\nu,\text{det}}}$ 
    \item Use the corrected maps to recompute a new frequency-averaged map and repeat steps 3–4. The process continues iteratively until convergence.
\end{enumerate}
We define one iteration of the algorithm as a pass through all detectors in a frequency band, where for each detector a new estimate of $\tilde{\theta}_\text{pol}^{\nu,\text{det}}$ is computed and applied. Between three and five iterations are typically required for the algorithm to converge to a self-consistent solution.\\
The procedure is iterative because the frequency-averaged map used as a reference in step 3 initially includes uncorrected detector miscalibrations. By progressively refining the calibration angles and re-estimating the averaged map from the corrected detector maps, the algorithm converges towards a solution where all detector maps are aligned. This iteration ensures that the final frequency-band map is built from consistently calibrated data and not biased by the initial misalignments.\\
The algorithm presented makes no assumption about the emission sources measured by the detectors and works with any foreground model. The presence of foregrounds in the sky maps, which normally is a source of contamination to be removed, improves the performance of the algorithm by increasing the signal-to-noise ratio of the detector maps.\\
The problem of absolute calibration, which is not addressed by the cross-calibration pipeline proposed, can be solved either by performing on-the-ground calibration, via $EB$ nulling approaches like the one presented in \cite{Krachmalnicoff_EBnulling}, or by using known astrophysical sources. When combined with cosmic-variance-limited $EB$ nulling approaches, the cross-calibration pipeline enables an absolute calibration of each detector's polarisation angle constrained only by cosmic variance and instrumental noise.
\subsection{Impact of residual miscalibrations on the tensor-to-scalar ratio}
\label{subsec: Residual miscalib}
The aim of the proposed cross-calibration algorithm is to reduce systematic errors in the determination of the tensor-to-scalar ratio, $r$. We assess if the precision achieved by the pipeline meets the requirements of a modern space mission. We compare $r$ estimated from ideally-calibrated maps and from maps calibrated with an uncertainty matching the performance of the cross-calibration algorithm. The bias introduced on $r$ by the miscalibration is then compared with \textit{LiteBIRD}'s budget for each systematic effect, $6.5\times10^{-6}$ \cite{PTEP}.\\
The pipeline used to estimate the bias on $r$ is similar to the one presented in \cite{Tommaso} and \cite{Florie}, and is as follows:\\
\begin{enumerate}
    \item Simulate a set of single-detector ideally-calibrated maps $m^{\text{ideal}}_{\nu,\text{det}}$ composed of a CMB realisation $m_\text{CMB}$, foregrounds and noise as described in \cref{subsec: Simulated Sky-Maps}.
    \item Apply a random miscalibration to each detector as described in \cref{subsec: Relative miscalibrations} using $\sigma_\text{cc}(\theta_\text{pol}^{\nu})$ equal to the calibration uncertainty achieved by the cross-calibration algorithm. The result is a new set of maps $m^{\text{obs}}_{\nu,\text{det}}$.
    \item Combine single detector maps into two sets of 22 frequency channel maps: $m^{\text{ideal}}_{\nu}$ and $m^{\text{obs}}_{\nu}$.
    \item Apply component separation to both sets of maps, obtaining two CMB $B$-mode maps: $m^{\text{ideal}}_{\text{CMB}}$ and $m^{\text{obs}}_{\text{CMB}}$, 
    \item Subtract the original CMB realisation  $m_\text{CMB}$ from the two maps to calculate a residual map. Then apply a \textit{Planck} 70\% mask \cite{Planck_PowerSpectra} to exclude the Galactic plane.
    \item For each simulation compute the angular power spectra of the residuals: $C_\ell^{\text{res},\text{ideal}}$ and $C_\ell^{\text{res},\text{obs}}$.
    \item Derive the best-fit value of $r$ by maximizing the following likelihood both in the ideal and the miscalibrated case:
    \begin{multline}
        -\ln \mathcal{L}(C_\ell^\text{obs}|r)=\sum_\ell \frac{2\ell+1}{2}f_\text{sky} \left[\frac{C_\ell^\text{obs}}{C_\ell^\text{th}(r)} +\ln\left(C_\ell^\text{th}(r)\right) -\frac{2\ell-1}{2\ell+1}\ln\left(C_\ell^\text{obs}\right)\right]
    \end{multline}
    where the observed $B$-mode spectrum is given by $C_\ell^\text{obs}=C_\ell^\text{res}+C_\ell^\text{lensing}$, and the theoretical one is computed as $C_\ell^\text{th}(r)=r\, C_\ell^{\text{GW},r=1}+C_\ell^\text{lensing}+\langle C_\ell^\text{res}\rangle_\text{sims}$. The term $\langle C_\ell^\text{res}\rangle_\text{sims}$ is a template of noise and foreground residuals after component separation, which is obtained by averaging $C_\ell^\text{res, ideal}$ over multiple noise and CMB realisations.
    \item Repeat the entire procedure for multiple CMB and noise realisations to construct a histogram of $\delta_r = r_\text{obs} - r_\text{ideal}$. The RMS of this distribution quantifies the combined effect of a systematic bias and an increased variance on $r$ due to the miscalibration. This is then compared with the \textit{LiteBIRD} systematic error budget.
\end{enumerate}
We perform this analysis using both \texttt{FgBuster}\footnote{https://github.com/fgbuster/fgbuster} \cite{FgBuster} and \texttt{HILC} \cite{ILC} as component-separation pipelines, representing parametric and blind approaches respectively. This allows us to assess whether the achieved calibration uncertainty is sufficient to meet the requirements of modern CMB experiments using different separation strategies.
\section{Results}
\label{sec: Results}
\subsection{Cross-Calibration uncertainty}
\begin{figure}[t!]
\centering
 \includegraphics[height=.8\textheight]{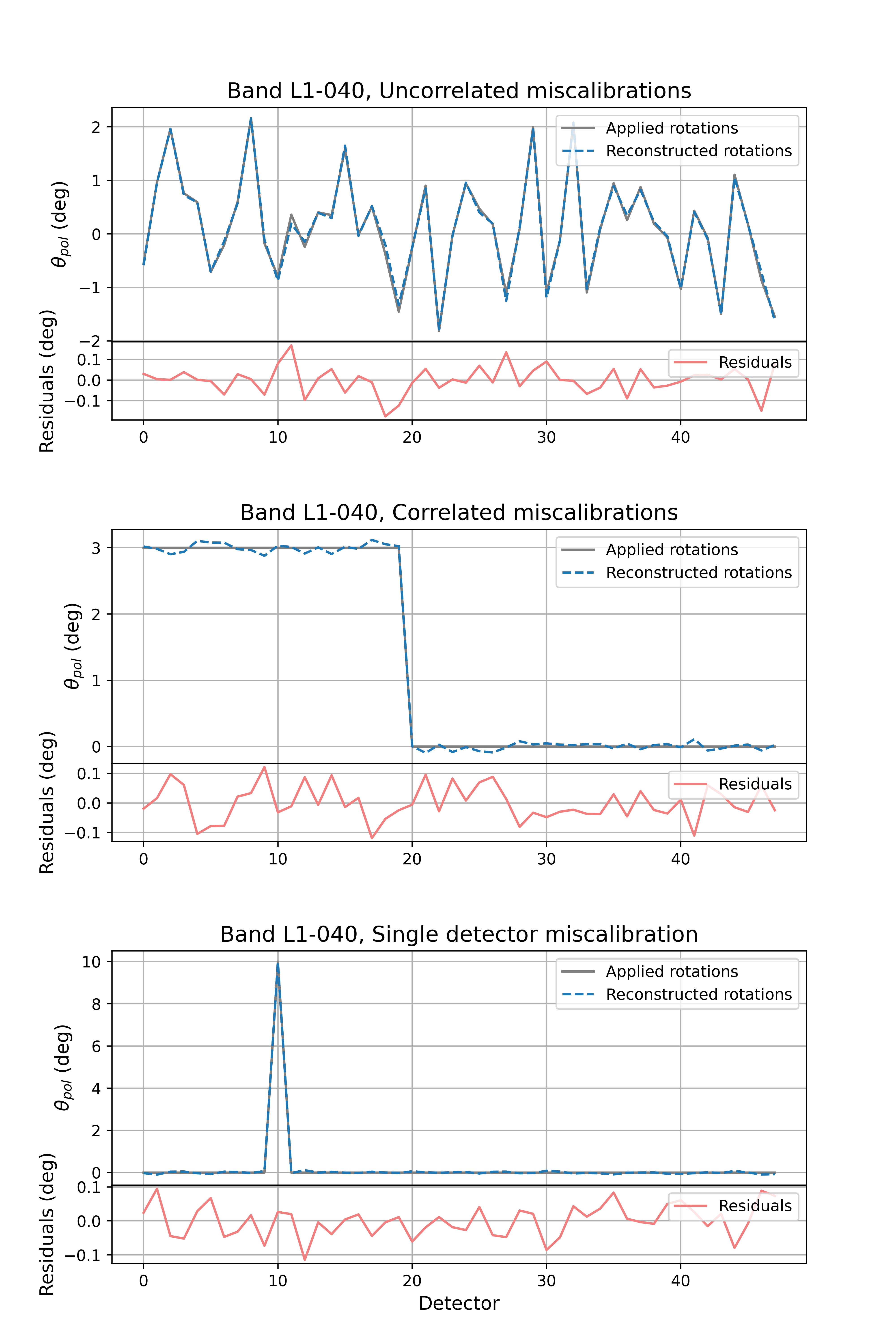}
\caption{Application of the cross-calibration algorithm to the LFT-$40\,\giga\hertz$ band of \textit{LiteBIRD}. The plots show the applied polarisation angle miscalibrations (gray), the recovered miscalibrations (blue dashed), and residuals (red) in three scenarios (top, middle and bottom panel). 
The top panel shows our results for the case in which errors of individual detectors are Gaussian distributed with standard deviation $\sigma(\theta^{\nu}_\text{pol})=1\degree$. The middle panel displays the results obtained when the first 20 detectors have a common angle miscalibration of $3\degree$ while the rest of the detectors are perfectly calibrated. In the bottom panel we assign a miscalibration of $10\degree$ to one detector and no miscalibration to the other detectors.}
\label{fig: theta_recon}
\end{figure}
\begin{figure}[t!]
\centering
 \includegraphics[height=.7\textheight]{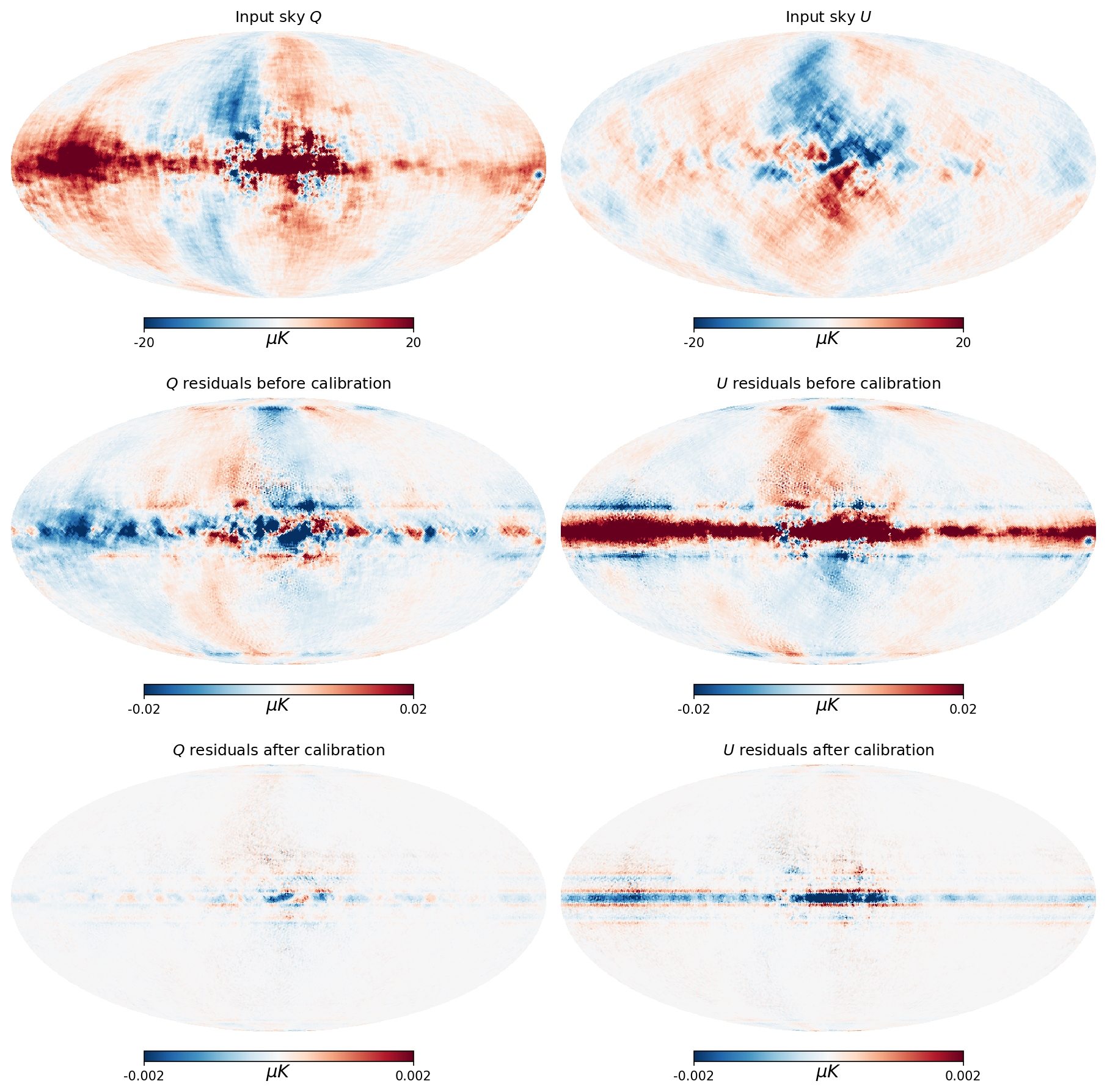}
\caption{$Q$ and $U$ sky maps of the LFT-40\,GHz channel. 
Top row: input sky maps. 
Middle row: residuals with respect to the input sky before cross-calibration. 
Bottom row: residuals with respect to the input sky after cross-calibration. 
To isolate the effect of the polarisation angle miscalibration, the noise 
contribution has been subtracted from the residual maps.
The miscalibration is drawn from a Gaussian distribution with standard 
deviation $\sigma(\theta^{\nu}_{\rm pol}) = 1^{\circ}$ and zero mean, 
consistent with the relative nature of the cross-calibration algorithm. 
After cross-calibration the residuals are suppressed by more than one 
order of magnitude, demonstrating the effectiveness of the algorithm at 
the map level. The horizontal stripes in the residual maps are introduced by \textit{LiteBIRD}'s scanning strategy. }
\label{fig: residual_maps}
\end{figure}
\begin{figure}[t]
\centering
\includegraphics[width=0.8\textwidth]{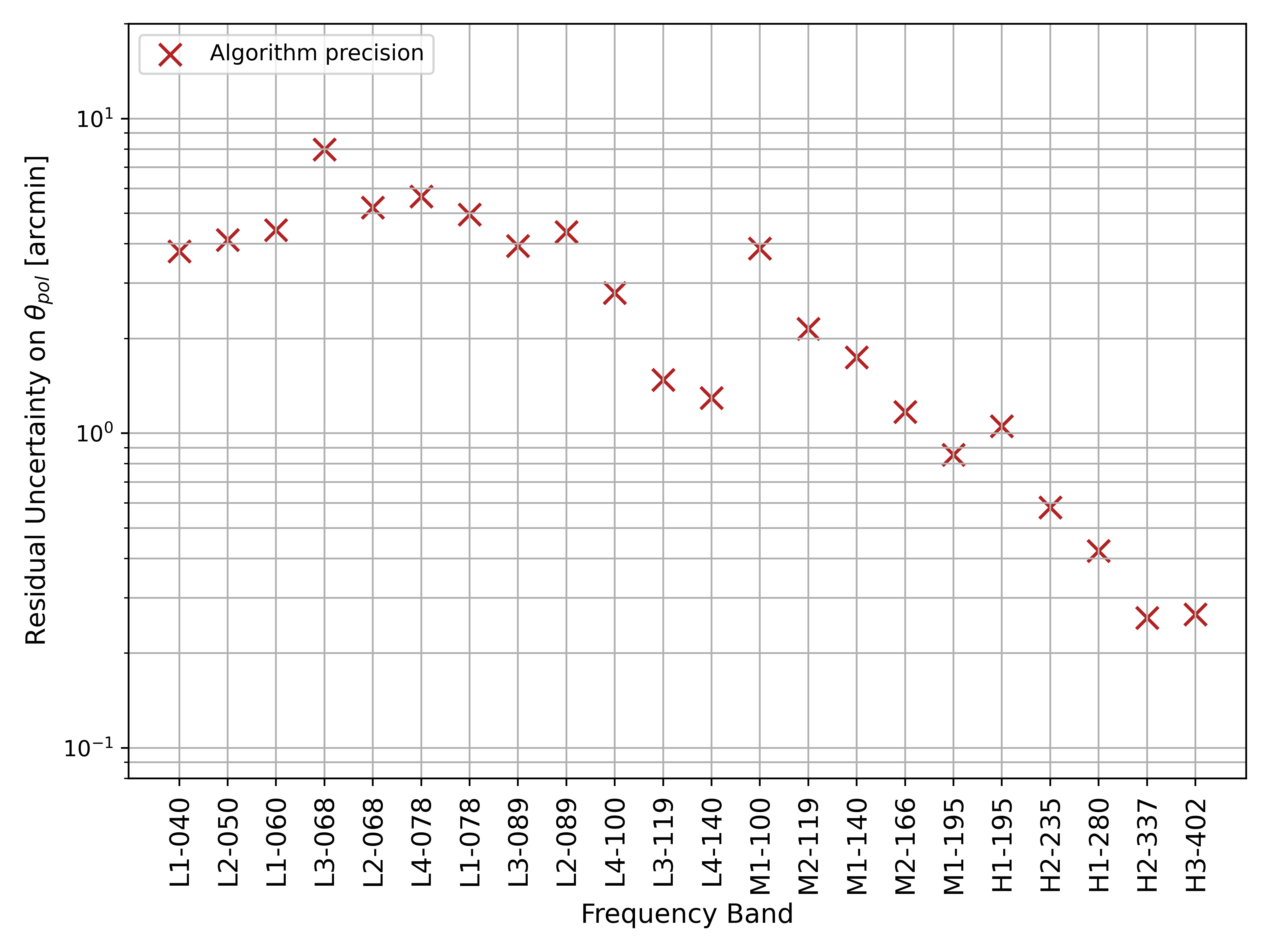}%
\caption{Residual polarisation angle uncertainty after cross-calibration for each channel of \textit{LiteBIRD} $\sigma_\text{cc}(\theta^\nu_\text{pol})$.}
\label{fig: residual_miscalib}
\end{figure}

We apply the cross-calibration algorithm described in \cref{subsec: Cross-calibration} to simulated \textit{LiteBIRD} maps that were generated according to the procedure detailed in \cref{subsec: Simulated Sky-Maps}. These maps have been injected with an initial polarisation angle uncertainty of $\sigma(\theta_\text{pol}^{\nu})=1 \degree$ across all frequency bands. The algorithm converges within 3–4 iterations, successfully identifying and correcting the polarisation miscalibration of each detector.\\
To assess the robustness of the calibration method against correlated miscalibrations and outliers, we investigate two additional scenarios. In the first scenario a subset of detectors is given a common miscalibration of $3\degree$, simulating the rotation of an entire focal plane wafer. In the second scenario a single detector is miscalibrated by $10\degree$, representing an isolated outlier. As shown in figure \ref{fig: theta_recon}, for the $40\, \giga \hertz$ channel the algorithm reliably recovers the true angles in all three scenarios, demonstrating robustness to both detector correlations and extreme outliers. The ability to identify and correct correlated miscalibrations can significantly relax the instrument's polarisation angle calibration requirements, since correlated errors are what primarily drives the need for stricter tolerances \cite{Patricio}. The effectiveness of the cross-calibration algorithm is further demonstrated 
at the map level in figure \ref{fig: residual_maps}, which shows the $Q$ and 
$U$ residuals before and after calibration for the LFT-40\,GHz channel. 
After cross-calibration the residuals are suppressed by more than one order of magnitude with respect to the miscalibrated maps.\\
To quantify the calibration uncertainty we compute the standard deviation of the residual miscalibrations over a large number of simulations.  As shown in figure \ref{fig: residual_miscalib}, this quantity varies across frequency bands, ranging from $0.2\,\text{arcmin}$ for high-frequency channels to $8\,\text{arcmin}$ for mid-frequency and low-frequency channels. The dominant factor influencing the calibration uncertainty is the signal-to-noise ratio (SNR) of the frequency channel. High-frequency channels, despite their higher noise levels, observe brighter foreground emission, which improve SNR and lead to better cross-calibration uncertainty. The resulting calibration uncertainties, denoted as $\sigma_\text{cc}(\theta_\text{pol}^{\nu})$ and displayed in \cref{tab: residual_miscalib}, are the values adopted in the following section to assess the impact of residual miscalibrations on the measurement of the tensor-to-scalar ratio.
\begin{table}[ht]
\centering
\begin{tabular}{cccc}
\hline
\textbf{Instrument} & \textbf{Freq.} & \textbf{Channel label} & \textbf{ Relative Cross-Calibration} \\
 &  &  & \textbf{Uncertainty $\sigma_\text{cc}(\theta^\nu_\text{pol})$} \\
&  [GHz] & & [arcmin] \\
\hline
LFT & 40  & LFT-40   & 3.8  \\
LFT & 50  & LFT-50   & 4.1 \\
LFT & 60  & LFT-60   & 4.4 \\
LFT & 68  & LFT-68a  & 8.0 \\
LFT & 68  & LFT-68b  & 5.2 \\
LFT & 78  & LFT-78a  & 5.7 \\
LFT & 78  & LFT-78b  & 5.0 \\
LFT & 89  & LFT-89a  & 3.9 \\
LFT & 89  & LFT-89b  & 4.4 \\
LFT & 100 & LFT-100  & 2.8 \\
LFT & 119 & LFT-119  & 1.5 \\
LFT & 140 & LFT-140  & 1.3 \\
MFT & 100 & MFT-100  & 3.9 \\
MFT & 119 & MFT-119  & 2.1 \\
MFT & 140 & MFT-140  & 1.7 \\
MFT & 166 & MFT-166  & 1.2 \\
MFT & 195 & MFT-195  & 0.9 \\
HFT & 195 & HFT-195  & 1.1 \\
HFT & 235 & HFT-235  & 0.6 \\
HFT & 280 & HFT-280  & 0.4 \\
HFT & 337 & HFT-337  & 0.3 \\
HFT & 402 & HFT-402  & 0.3 \\
\hline
\end{tabular}
\caption{Residual relative polarisation angle uncertainty after cross-calibration for each individual polarimeter of \textit{LiteBIRD}.}
\label{tab: residual_miscalib}
\end{table}

\subsection{Compliance with \textit{LiteBIRD}'s requirements}
\label{subsec: Res delta_r}
\begin{figure}[ht]
    \centering
    \includegraphics[width=\textwidth]{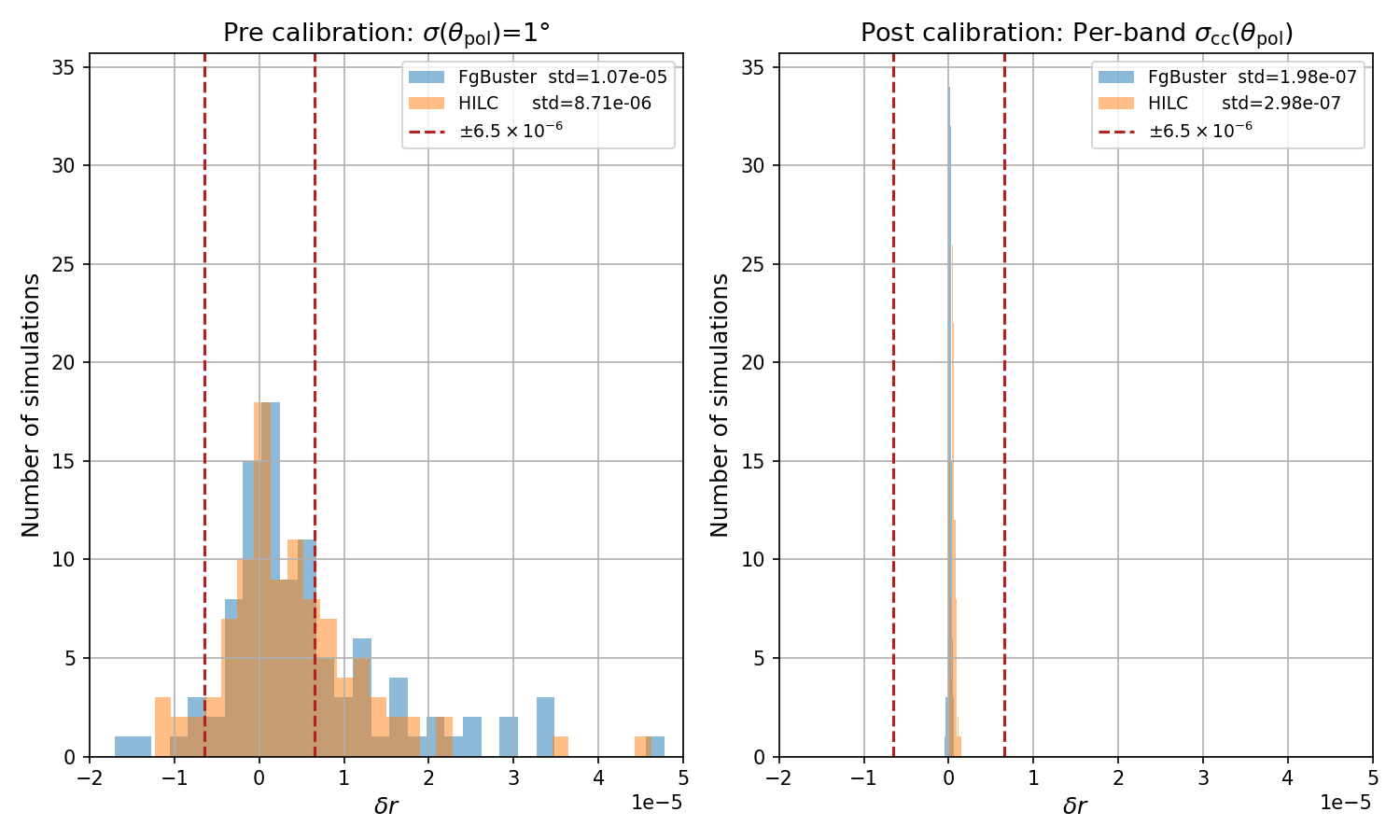}%
    \caption{Bias on $r$ for a pre-calibration uncertainty of $\sigma(\theta^\nu_\text{pol}) = 1\degree$ (left) and for the residual miscalibrations after cross-calibration (right). In both panels, the blue histogram is obtained using \texttt{FgBuster} and the orange one using \texttt{HILC}. The dashed red lines indicate the \textit{LiteBIRD} systematic error budget of $\pm 6.5\times10^{-6}$. Before calibration, the budget is exceeded in 36\% and 38\% of simulations for \texttt{FgBuster} and \texttt{HILC} respectively. After cross-calibration, all simulations remain well within the budget for both component separation methods.}
    \label{fig: Delta_r}
\end{figure}

A detailed study of the requirements on polarisation angle calibration to fulfil \textit{LiteBIRD}'s scientific goal was carried out in \cite{Patricio}. That paper also presented results for the case of fully uncorrelated miscalibrations between all detectors, as is the case for the residuals left by our cross-calibration algorithm. The reported requirements on polarisation angle uncertainty are at least an order of magnitude more relaxed than the precision achieved by our cross-calibration algorithm, suggesting that this method, if coupled with an appropriate way to perform absolute calibration, is more than sufficient for \textit{LiteBIRD}'s needs. However, the analysis in \cite{Patricio} assumes perfect component separation despite the presence of systematics. This assumption decouples miscalibrations from the additional component-separation residuals that are introduced by the miscalibration. In other words, it neglects the fact that larger miscalibrations worsen component separation, increasing foregrounds to CMB leakage, thereby underestimating the strictness of the requirements. To provide a more realistic assessment, we use the pipeline presented in \cref{subsec: Residual miscalib}, which incorporates miscalibrations directly into the component separation and $r$-estimation pipeline. This approach yields more accurate values for the bias $\delta_r$ introduced by miscalibrations.\\
In order to measure the tensor-to-scalar ratio with a precision of $\sigma_r\simeq 10^{-3}$, the \textit{LiteBIRD} collaboration has assigned a maximum allowable contribution of $\delta_r=6.5\times10^{-6}$ from each systematic effect. Using the procedure described in \cref{subsec: Residual miscalib} we evaluate two cases: first, whether a polarisation angle uncertainty of $\sigma(\theta^\nu_\text{pol}) = 1\degree$ is already sufficient to exceed the budget \footnote{$\sigma(\theta^\nu_\text{pol}) = 1\degree$ is consistent with the typical mechanical misalignment of individual bolometers \cite{deBernardis2016}. Note that the requirements derived in \cite{Patricio} are  $> 1\degree$, suggesting that $\sigma(\theta^\nu_\text{pol}) = 1\degree$ should be sufficient. However, as discussed above, that analysis assumes perfect component separation, which underestimates the impact of miscalibrations on $\delta_r$. Our more realistic assessment shows that even $\sigma(\theta^\nu_\text{pol}) = 1\degree$ can be sufficient to violate the budget.}; second, whether the residual miscalibrations after cross-calibration $\sigma_\text{cc}(\theta^\nu_\text{pol})$ remain within the budget.\\
For each case, we generate two sets of \textit{LiteBIRD} maps with the same noise and CMB realisation: one ideally-calibrated and the other with a polarisation angle uncertainty of either $1\degree$ or $\sigma_\text{cc}(\theta_\text{pol}^{\nu})$, depending on the case. We then apply the pipeline from \cref{subsec: Residual miscalib} to both sets of maps, using first \texttt{FgBuster} and then \texttt{HILC} for component separation. This allows us to estimate the bias on $r$ introduced by the residual miscalibrations using both a parametric and a blind component-separation method.\\
After 100 simulations, we obtain the distributions for $\delta_r$ depicted in figure \ref{fig: Delta_r}. In the uncalibrated case with $\sigma(\theta^\nu_\text{pol}) = 1\degree$ (left), the budget is exceeded in 36\% of simulations when using \texttt{FgBuster} and 38\% when using \texttt{HILC} demonstrating that a pre-calibration uncertainty of $1\degree$ is insufficient to meet \textit{LiteBIRD}'s requirements. After cross-calibration (right), the magnitude of $\delta_r$ remains well within the \textit{LiteBIRD} requirement of $6.5\times10^{-6}$ in all simulations for both component separation methods.
\section{Conclusions}
\label{sec: Conclusions}
In this paper we have presented a relative cross-calibration method that performs relative calibration of detectors within the same frequency band of a CMB experiment. The algorithm can be applied to any systematic that can be corrected via the re-processing of the instrument's data at the map level, such as gain, polarisation efficiency, or polarisation angle. The pipeline is designed to perform relative calibration only, the problem of absolute calibration can be addressed either through ground-based measurements, through $E$-$B$ nulling techniques such as the one presented in \cite{Krachmalnicoff_EBnulling} or using known astrophysical sources.\\
We demonstrated the effectiveness of the method by applying it to the calibration of the polarisation angle in \textit{LiteBIRD} simulated maps. The cross-calibration algorithm performs robustly in the presence of random miscalibrations, correlated miscalibrations, and outlier detectors with very large miscalibrations. After correction, the residual polarisation angle uncertainties range between $0.2\,\text{arcmin}$ and $8\,\text{arcmin}$ depending on the frequency band, and show no significant residual correlations.\\
We showed that these residual relative miscalibrations induce a bias on the tensor-to-scalar ratio $r$ that is well below \textit{LiteBIRD}'s systematic error budget of $6.5\times10^{-6}$. This result holds for both parametric and blind component separation methods, making the algorithm suitable for use by modern CMB experiments.\\
Future work will aim to determine the performance of the algorithm in calibrating different detector properties such as their gain or polarisation efficiency.\\
Overall, the proposed cross-calibration method provides a flexible and robust tool to meet the stringent calibration requirements of future CMB experiments.

\acknowledgments
This work has also received funding by the European Union’s Horizon 2020 research and innovation program under grant agreement no. 101007633 CMB-Inflate.
\vspace{1mm}\\
\textit{LiteBIRD} (phase A) activities are supported by the following funding sources: ISAS/JAXA, MEXT, JSPS, KEK (Japan); CSA (Canada); CNES, CNRS, CEA (France);
DFG (Germany); ASI, INFN, INAF (Italy); RCN (Norway); MCIN/AEI, CDTI (Spain); SNSA, SRC (Sweden); UKSA (UK); and NASA, DOE (USA).
\vspace{1mm}\\
We acknowledge support from the COSMOS network through the ASI (Italian Space Agency) Grants 2016-24-H.0 and 2016-24-H.1-2018, as well as 2020-9-HH.0 (participation in \textit{LiteBIRD}, phase A).




\bibliographystyle{JHEP}
\bibliography{biblio}
\end{document}

%% file: authors_affiliations_LB_2025.08_general_jcap.tex
\author[1,2]{A.\,Novelli,}
\author[2,3]{F.\,Piacentini,}
\author[4,2]{S.\,Micheli,}
\author[5]{E.\,Allys,}
\author[6]{A.\,Anand,}
\author[7]{J.\,Aumont,}
\author[7]{A.\,J.\,Banday,}
\author[8]{R.\,B.\,Barreiro,}
\author[9,10,11]{N.\,Bartolo,}
\author[12]{S.\,Basak,}
% ...
\affiliation[1]{University of Milano Bicocca, Physics Department, p.zza della Scienza, 3, 20126 Milan, Italy}
\affiliation[2]{Dipartimento di Fisica, Università La Sapienza, P. le A. Moro 2, Roma, Italy}
\affiliation[3]{INFN Sezione di Roma, P.le A. Moro 2, 00185 Roma, Italy}
\affiliation[4]{Université Grenoble Alpes, CNRS, LPSC-IN2P3, 53, avenue des Martyrs, 38000 Grenoble, France}
\affiliation[5]{Laboratoire de Physique de l’École Normale Supérieure, ENS, Université PSL, CNRS, Sorbonne Université, Université de Paris, 75005 Paris, France}
\affiliation[6]{Dipartimento di Fisica, Università di Roma Tor Vergata, Via della Ricerca Scientifica, 1, 00133, Roma, Italy}
\affiliation[7]{IRAP, Université de Toulouse, CNRS, CNES, UPS, Toulouse, France}
\affiliation[8]{Instituto de Fisica de Cantabria (IFCA, CSIC-UC), Avenida los Castros SN, 39005, Santander, Spain}
\affiliation[9]{Dipartimento di Fisica e Astronomia “G. Galilei”, Università degli Studi di Padova, via Marzolo 8, I-35131 Padova, Italy}
\affiliation[10]{INFN Sezione di Padova, via Marzolo 8, I-35131, Padova, Italy}
\affiliation[11]{INAF, Osservatorio Astronomico di Padova, Vicolo dell’Osservatorio 5, I-35122, Padova, Italy}
\affiliation[12]{School of Physics, Indian Institute of Science Education and Research Thiruvananthapuram, Maruthamala PO, Vithura, Thiruvananthapuram 695551, Kerala, India}
% ...
